# Crystal Structure and Physical Properties of $Mg_6Cu_{16}Si_7$-type $M_6Ni_{16}Si_7$, for M = Mg, Sc, Ti, Nb, and Ta

K. L. Holman[1], E. Morosan[1], P. A. Casey[2], Lu Li[2], N.P. Ong[2], T. Klimczuk[3,4], C. Felser[5] and R. J. Cava[1]

[1]Department of Chemistry, Princeton University, Princeton, NJ, 08540 USA
[2]Department of Physics, Princeton University, Princeton, NJ, 08540 USA
[3] Faculty of Applied Physics and Mathematics, Gdansk University of Technology, Narutowicza 11/12, 80-952 Gdansk, Poland
[4] Los Alamos National Laboratory, Los Alamos, NM 87545, USA
[5]Johannes Gutenberg-Universität, Institut für Anorganische Chemie und Analytische Chemie, Staudingerweg 9, 55128 Mainz, Germany

**Abstract**

Five compounds were investigated for magnetic character and superconductivity, all with non-magnetic nickel and band structures containing flat bands and steep bands. The syntheses and crystal structures, refined by powder X-ray diffraction, are reported for $M_6Ni_{16}Si_7$, where M = Mg, Sc, Ti, Nb, and Ta. All compounds form in the $Mg_6Cu_{16}Si_7$ structure type. Resistance measurements are also reported on $M_6Ni_{16}Si_7$ (M = Mg, Sc, Ti, and Nb) down to 0.3 K, with all four showing metallic conductivity. No superconductivity is observed. Magnetization measurements for all compounds reveal essentially temperature independent paramagnetism, with a tendency toward more enhanced low temperature paramagnetism for the cases of $Mg_6Ni_{16}Si_7$ and $Sc_6Ni_{16}Si_7$.

Keywords: A. intermetallic compounds, metals B. chemical synthesis, C. X-ray diffraction, D. superconductivity

**Introduction**

The prediction of superconductivity in a new material is generally limited to finding compounds with characteristics similar to known superconductors. Two classes of known superconducting compounds, $MgCNi_3$ [1] and $LnNi_2B_2C$ [2], are intermetallics based on the late *3d* transition metal Ni. Generally, nickel compounds are magnetic. However, when Ni *3d* states are filled or nearly filled by electrons donated from more electropositive elements, as in $MgCNi_3$ and $LnNi_2B_2C$, compounds can show non-magnetic character. In $MgCNi_3$ and $LnNi_2B_2C$, magnetic character is suppressed, and superconductivity occurs. Because few compounds with nearly filled Ni *3d* states have been characterized, it is unclear whether superconductivity is a general phenomenon in this class. Thus, it is important to investigate compounds with these crystal chemical characteristics to determine whether superconductivity is a general phenomenon.

In addition to the suppression of the magnetic character, other unifying characteristics are important to identify in the search for superconductors. In one study [3] a large number of known superconductors were investigated, using basic LMTO calculations, for distinctive band structures. Every superconductor possessed a band structure with both flat bands and steep bands. This is known as the 'flat band/steep band' scenario [3]. Although the existence of both flat and steep bands is not in itself enough to predict superconductivity, some consider it a necessary condition [4].

A family of compounds with promising band structures and crystal chemistries is based on the $Mg_6Cu_{16}Si_7$ structure type. The crystal structure is complex due to the large number and types of atoms involved. However, the structure is easily described (see below) in terms of simple intermetallic polyhedra and can sometimes display interesting structural variations. One variation, an incommensurate structure, has been observed via electron diffraction in rapidly quenched $V_6Ni_{16}Si_7$[5,6]. Some members of this family of compounds were synthesized early on [1,7,8,9]; many have not been tested for low temperature magnetic properties or superconductivity [1,7,9]. The chemical variability of the structure type allows for the synthesis of $M_6Ni_{16}Si_7$, where M=Mg, Sc, Ti, Nb, and Ta resulting in a valence electron difference over the series of up to 3 electrons per M atom (between Mg and Nb or Ta). This yields a maximum change over the synthesized series in electron count of 1.125 electrons per Ni. The variability in valence electron count within the unit cell within a rigid band scenario moves the Fermi energy substantially. Here we report a study of the crystal structures and physical properties of $M_6Ni_{16}Si_7$ compounds (M = Mg, Sc, Ti, Nb, and Ta), in order to test for the degree of magnetic suppression and for superconductivity

**Experimental**

The magnesium containing compound was prepared by combining stoichiometric amounts of Ni (Alfa, nanopowder, 98%) and Si (Alfa, submicron particles, 98%) by grinding and then combining with 20% excess Mg and pressing into a small pellet. The 20% excess was employed to counter the loss of Mg during heating [10]. The pellet was wrapped in Ta foil and heated under flowing 5 % $H_2$/ 95 % Ar for one hour at 500 °C and one hour at 900 °C. It was then cooled, ground, pressed again with 20 % more Mg and reheated for one more hour at 900 °C and bench cooled. Compounds without magnesium were made by arc melting. $M_6Ni_{16}Si_7$, where M = Sc, Ti, Nb, and Ta where made using Ni powder (Johnson Matthey, 99.9%), Si chunks (Alfa, 99.9999%) and Sc chunks (Alfa, 99.9%), Ti powder (Johnson Matthey, 99.9%), Nb powder (Cerac, 99.8%) or Ta powder (Alfa, 99%), respectively. All arc melted samples were pressed into pellets before melting. Each pellet was melted three times; weight loss, if any, was less than two percent. To eliminate impurities seen in powder X-ray diffraction data, the $Ta_6Ni_{16}Si_7$ and $Nb_6Ni_{16}Si_7$ samples were annealed in sealed quartz tubes at 900 °C for four days.

The crystal structures were characterized by X-ray powder diffraction, employing a Bruker D8 diffractometer with Cu Kα radiation with a graphite diffracted beam monochromator. The software TOPAS 2.1 (Bruker AXS) was used for Rietveld structure refinements. The published crystal structure of $Mg_6Cu_{16}Si_7$ was employed as a starting structural model [7]. Magnetization measurements were performed in a Quantum Design PPMS system. For $Mg_6Ni_{16}Si_7$ samples, field dependent magnetization measurements indicated the presence of trace amounts of Ni metal impurity (about 1 part per 1000). Therefore, the susceptibility of $Mg_6Ni_{16}Si_7$ was determined from the difference in measured magnetizations at applied fields of 5 T and 2 T ($\chi = \Delta M/\Delta H$), where the M vs. H relationships were linear. This procedure was performed in the temperature range 5 K to 250 K, with a temperature step of 5 K. This technique has been used previously to measure the intrinsic susceptibility of $MgCNi_3$, which also typically contains a trace amount of ferromagnetic elemental Ni impurity [1]. Resistance measurements were made above 2 K on bars fabricated from as synthesized materials in a Quantum Design PPMS using the four probe method with silver paint contacts and platinum wires. Below 2 K, four probe AC resistivity was measured at low frequency in a Janis He-3 crystat.

Self-consistent electronic structure calculations were performed using the TB-LMTO-ASA (tight binding-linearized muffin tin orbital-atomic sphere approximation) codes developed by the Andersen group [11]. A detailed description of the local density approximation, the reciprocal space integration, and the procedure by which self-consistency is achieved, can be found elsewhere [12, 13, 14, 15]. 5000 irreducible k-points were used.

**Results**

Preliminary calculations of the band structure and density of states for all five compounds by the TB-LMTO-ASA method indicated promising results. An example of these calculations for $Mg_6Ni_{16}Si_7$ is shown in Figure 1. Around the Fermi energy, both flat and steep bands are present. The flat bands have a strong nickel contribution and lead to the expected density of states peak at the Fermi energy. This peak at the Fermi energy is observed in all compounds independent of the number of valence electrons. The nickel states are nearly filled in all five compounds and found mainly between -4 and -1 eV. Based on this information, the decision was made to attempt the synthesis and to test five materials for superconductivity.

All five compounds were found to be isostructural to $Mg_6Cu_{16}Si_7$, a cubic compound with a cell parameter of 11.65 Å and space group *Fm3m* [7]. An example of the observed X-ray data, the calculated powder diffraction pattern, the difference between the calculated model and the observed data, and positions of expected peaks is presented in Figure 2 for $Mg_6Ni_{16}Si_7$. As shown in Figure 1, the agreements between the model and the data were excellent, with $\chi^2 \approx 1.1$. The refined structural parameters for all five compounds are presented in Table 1; each unit cell contains four formula units. There are only three atomic position parameters to be determined in the structural refinements of $M_6Ni_{16}Si_7$: the M atom in position (x,0,0), with x ≈ 0.2, and two different Ni atoms in the (x,x,x) position, with x ≈ 0.17 and x ≈ 0.38. The Si are on the fixed positions (½, ½, ½) and (0, ¼, ¼). The freely refined atomic site occupancies were within three standard deviations of one indicating that all five compounds are stoichiometric. Therefore, the site occupancies were fixed at one for the final refinement, which included isotropic temperature factors.

The crystal structure of these compounds, using $Mg_6Ni_{16}Si_7$ as an example, can be described as follows. The foundation of the structure is an edge sharing network of $Ni_8Si_{10}$ supertetrahedra with hollow Mg octahedra in the interstices. The heart of each supertetrahedron is an 8 atom Ni cluster formed from two interpenetrating Ni tetrahedra (Fig. 3a). Each $Ni_8$ cluster is surrounded by a silicon octahedron (Fig. **3**b): four of the Ni atoms sit within the four tetrahedrally opposing faces of the Si octahedron, and the other 4 Ni atoms cap the remaining 4 faces of the Si octahedron. Finally, the $Ni_8Si_{10}$ supertetrahedron is formed when four additional Si cap the Si octahedron above the four Ni that lie above the faces of the internal Si octahedron (Fig. **3**c). Eight of these $Ni_8Si_{10}$ supertetrahedra, each approximately 8 Å on an edge, are arranged in each unit cell in a fully edge sharing geometry. The edge sharing array of $Ni_8Si_{10}$ supertetrahedra creates interstitial pockets. The interstitial pockets contain distinct octahedra consisting of 6 Mg atoms at the vertices of a hollow octahedron (one of which is shown in Fig. **3**d). The complete unit cell is composed of a face centered cubic pattern of Mg octahedra combined with eight edge sharing supertetrahedra (Fig. 4). In simplest terms, the structure, consisting of an FCC arrangement of $Mg_6$ octahedra with $Ni_8Si_{10}$ supertetrahedra in all the tetrahedral holes, can be considered as the $CaF_2$ fluorite-type.

The temperature dependent magnetic susceptibilities are shown in the main panel of Figure 5. All materials have a relatively small magnetic susceptibility, which is essentially independent of temperature between 5 K and 250 K. Magnetization versus field data for $Mg_6Ni_{16}Si_7$ and $Sc_6Ni_{16}Si_7$ are shown in the inset to Figure 5. $Mg_6Ni_{16}Si_7$ contains approximately 0.1% elemental Ni impurity that is evident by the nonlinearity of the magnetization versus field data. Therefore, magnetization versus temperature values were calculated as described above. The remaining four compositions show strictly linear M versus H behavior at all temperatures in the field range 0-5 T, as illustrated by $Sc_6Ni_{16}Si_7$ in the inset of Figure 5. Though the magnitudes of the susceptibilities do not change significantly across the series, the data for $Mg_6Ni_{16}Si_7$ and $Sc_6Ni_{16}Si_7$ do suggest the tendency towards increased paramagnetism at low temperatures. Thus, although weak magnetism is clearly not present in these materials, it may be beginning to emerge for the lower electron count variants studied.

Resistance measurements for $Mg_6Ni_{16}Si_7$, $Sc_6Ni_{16}Si_7$, $Ti_6Ni_{16}Si_7$, and $Nb_6Ni_{16}Si_7$ all show metallic behavior (Fig. 6). While $Nb_6Ni_{16}Si_7$ shows little dependence on temperature, $Mg_6Ni_{16}Si_7$ has more metallic-like behavior, with a low temperature resistivity of about 0.14 mΩ-cm, compared with 0.43 mΩ-cm for $Nb_6Ni_{16}Si_7$. $Ti_6Ni_{16}Si_7$ is a very poor metal, displaying a high, temperature-independent resistivity of about 1.5 mΩ-cm. $Sc_6Ni_{16}Si_7$ is the best metal of the group, with a resistivity at 0.4 K of approximately 0.04 mΩ-cm. No magnetoresistance was observed in fields up to 4 Tesla, and no superconductivity was observed.

**Conclusions**

Five compounds, $M_6Ni_{16}Si_7$ with M= Mg, Sc, Ti, Nb, and Ta, were selected for investigation due to their characteristics in common with known superconductors. All five compounds have non-magnetic nickel. Our preliminary electronic structure calculations using the LMTO method suggest that both flat bands and steep bands are present at the Fermi energy. Investigation of the magnetizations yielded largely temperature independent susceptibilities with the slight exception of $Mg_6Ni_{16}Si_7$ and $Sc_6Ni_{16}Si_7$, which both showed low temperature paramagnetism. Measurement of the resistivities of $M_6Ni_{16}Si_7$ with M=Mg, Sc, Ti and Nb showed all to be metallic conductors. Superconductivity was not found in any of the investigated systems.

**Acknowlegements**

This research was supported by the National Science Foundation, grant DMR-0213706, and by the Department of Energy Division of basic Energy Sciences, grant DE-FG02-98ER45706.

Figure Captions

Fig. 1 A representation of the band structure of $Mg_6Ni_{16}Si_7$ shows both flat and steep bands as well as a peak in the density of states (DOS) at the Fermi energy.

Fig. 2 Powder X-ray diffraction of a single phase sample fitting the $Mg_6Ni_{16}Si_7$ structure type. $Mg_6Ni_{16}Si_7$ powder X-ray diffraction pattern measured from 10 to 90 degrees showing the measured (squares), calculated (red line) and difference (grey line) plots. The vertical lines indicate the position of calculated peaks.

Fig. 3 The $Mg_6Ni_{16}Si_7$ structure is constructed using a series of polyhedra. The building blocks for the crystal structure of $Mg_6Cu_{16}Si_7$-type $Mg_6Ni_{16}Si_7$ are shown; only portions of the unit cell are shown for clarity. (a) The $Ni_8$ cluster made from interpenetrating $Ni_4$ tetrahedra; (b) The manner in which a $Si_6$ octahedron surrounds each $Ni_8$ cluster: 4 Ni are in the faces of the $Si_6$ octahedron and 4 are above the faces; (c) 4 more Si are added to the cluster to create an $Ni_8Si_{10}$ supertetrahedron. A single supertetrahedron is shown; (d) A single $Mg_6$ octahedron cluster.

Fig. 4 The polyhedra from Figure 2 assembled into the full unit cell of $Mg_6Ni_{16}Si_7$. The full unit cell of $Mg_6Ni_{16}Si_7$, showing the $Mg_6$ octahedra fitting in the cavities in the edge-sharing array of $Ni_8Si_{10}$ supertetrahedra. The structure can be viewed as an FCC packing of $Mg_6$ octahedra with $Ni_8Si_{10}$ supertetrahedra in all the tetrahedral interstices – a cluster-based intermetallic fluorite ($CaF_2$) structure.

Fig. 5 $Sc_6Ni_{16}Si_7$ and $Mg_6Ni_{16}Si_7$ show weakly temperature dependent paramagnetism with other compositions being essentially temperature independent paramagnets. Temperature dependent magnetic susceptibilities for all 5 compounds. Inset: The field

dependent magnetization measurements at 5 K are shown for $Mg_6Ni_{16}Si_7$ and $Sc_6Ni_{16}Si_7$. In the range of 0 and 5T, the remaining three samples showed strictly linear M vs H behavior.

Fig. 6 The four compounds measured are metallic with $Sc_6Ni_{16}Si_7$ showing very low resistivities. The temperature dependent resistivities for $Ti_6Ni_{16}Si_7$, $Mg_6Ni_{16}Si_7$, $Nb_6Ni_{16}Si_7$ and $Sc_6Ni_{16}Si_7$ measured from 0.3 K to 300K are shown.

Table 1: Atomic coordinates and isotropic displacement parameters (Å$^2$) for M$_6$Ni$_{16}$Si$_7$, spacegroup Fm-3m (225)

|  | M=Mg | Sc | Ti | Nb | Ta |
|---|---|---|---|---|---|
| a$_0$ (Å)= | 11.3164(19) | 11.44419(86) | 11.2566(13) | 11.244(29) | 11.2168(14) |
| M (x,0,0) *(24 e)* x= | 0.20320(46) | 0.20599(65) | 0.20671(49) | 0.20045(29) | 0.20229(35) |
| *U*(eq)= | 0.0163(19) | -0.0016(77) | 0.0206(90) | 0.0089(14) | 0.0015(35) |
| Ni(1) (x,x,x) *(32 f)* x= | 0.16820(12) | 0.16807(32) | 0.16205(20) | 0.16792(24) | 0.16858(53) |
| *U*(eq)= | 0.02134(85) | -0.0032(76) | 0.0219(91) | 0.0058(19) | 0.0008(48) |
| Ni(2) (x,x,x) *(32 f)* x= | 0.38284(14) | 0.38082(31) | 0.39028(17) | 0.38326(23) | 0.38335(43) |
| *U*(eq)= | 0.01682(90) | 0.0008(80) | 0.0215(94) | 0.0070(20) | 0.012(5) |
| Si(1) (0,¼,¼) *(24 d)* *U*(eq)= | 0.0210(16) | 0.0135(82) | 0.0016(94) | 0.0013(28) | 0.0124(92) |
| Si(2) (½,½,½) *(4 b)* *U*(eq)= | 0.0117(34) | 0.014(14) | 0.0250(107) | 0.0152(79) | -0.001(23) |
| $\chi^2$= | 1.09 | 1.11 | 1.20 | 1.14 | 1.12 |

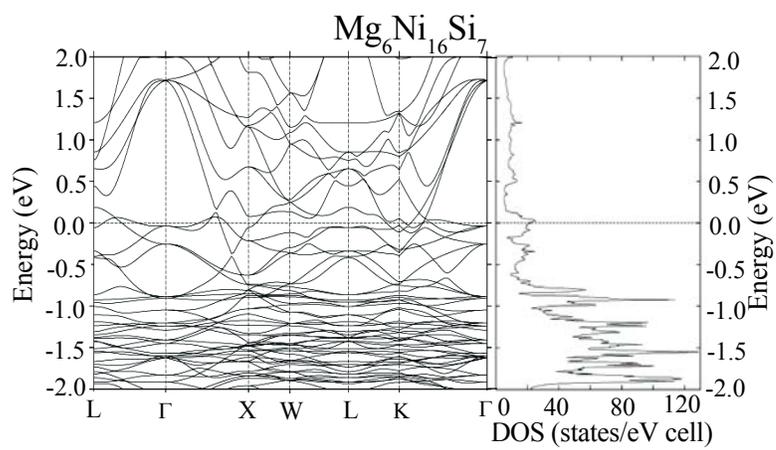

Figure **1**

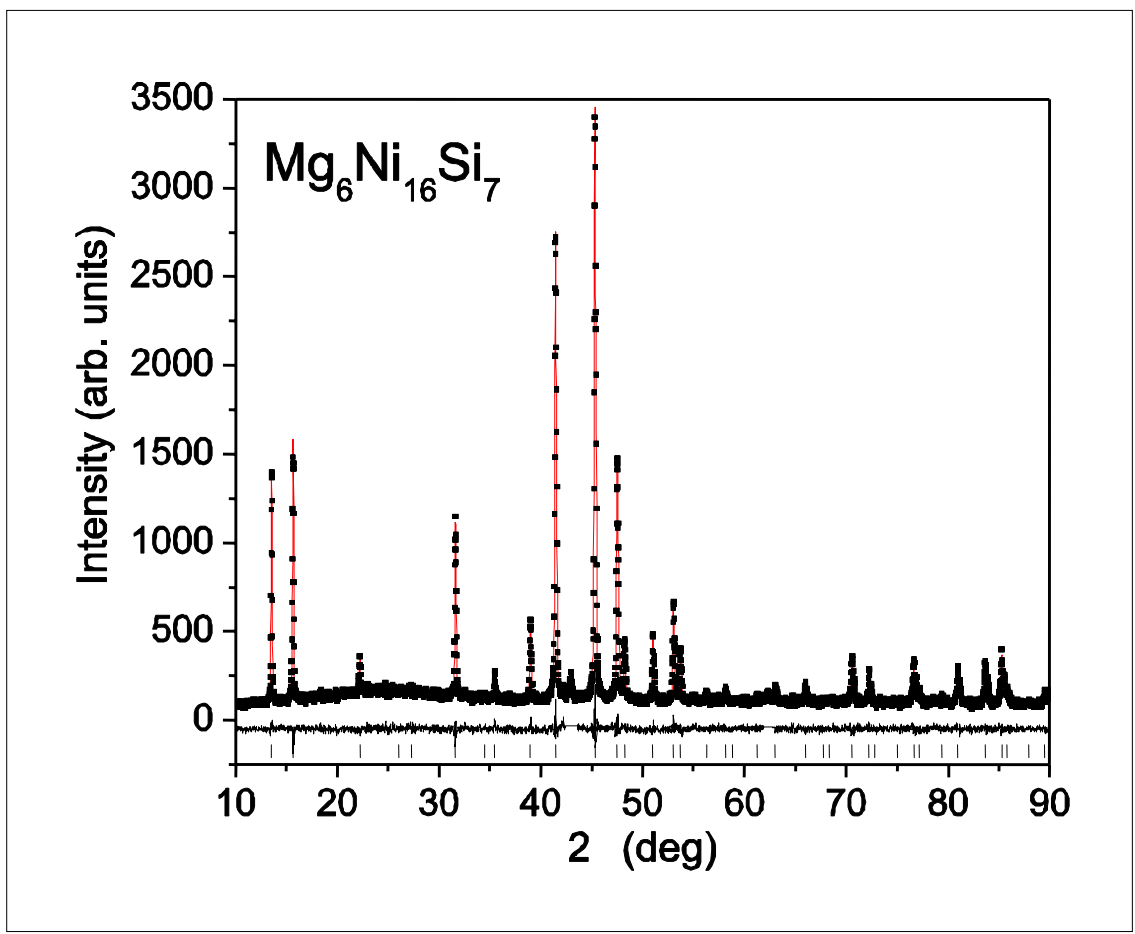

Figure **2**

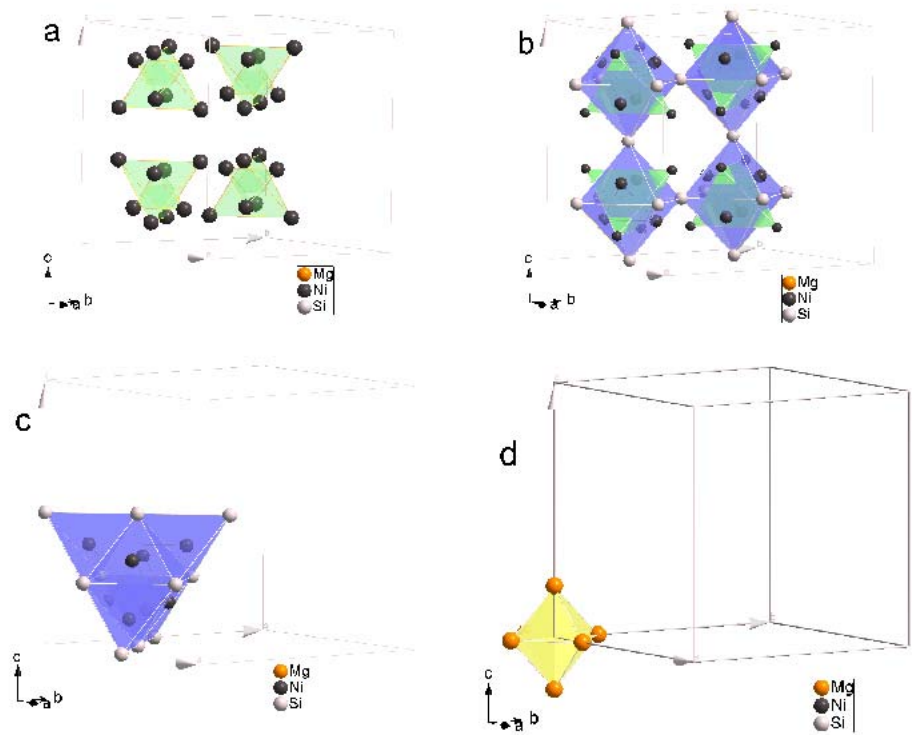

Figure **3**

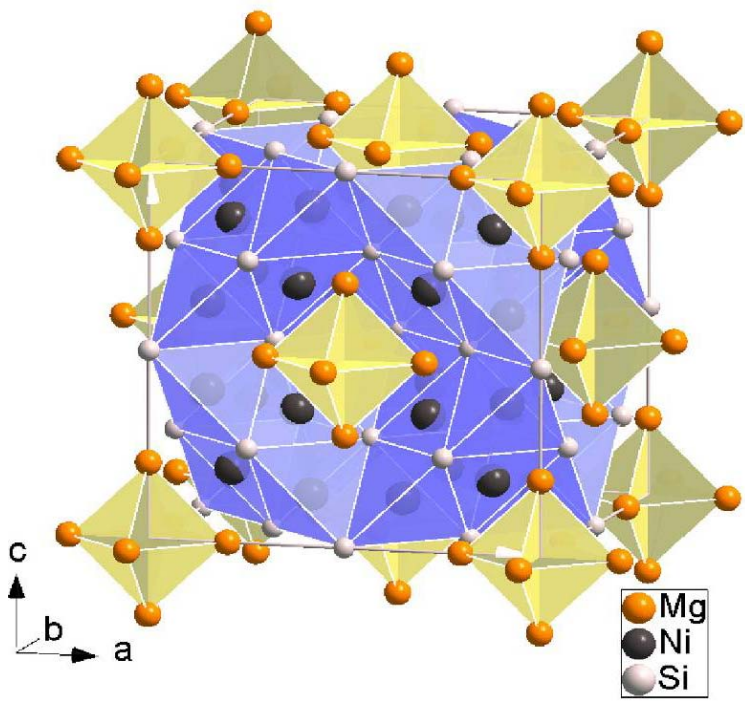

Figure **4**

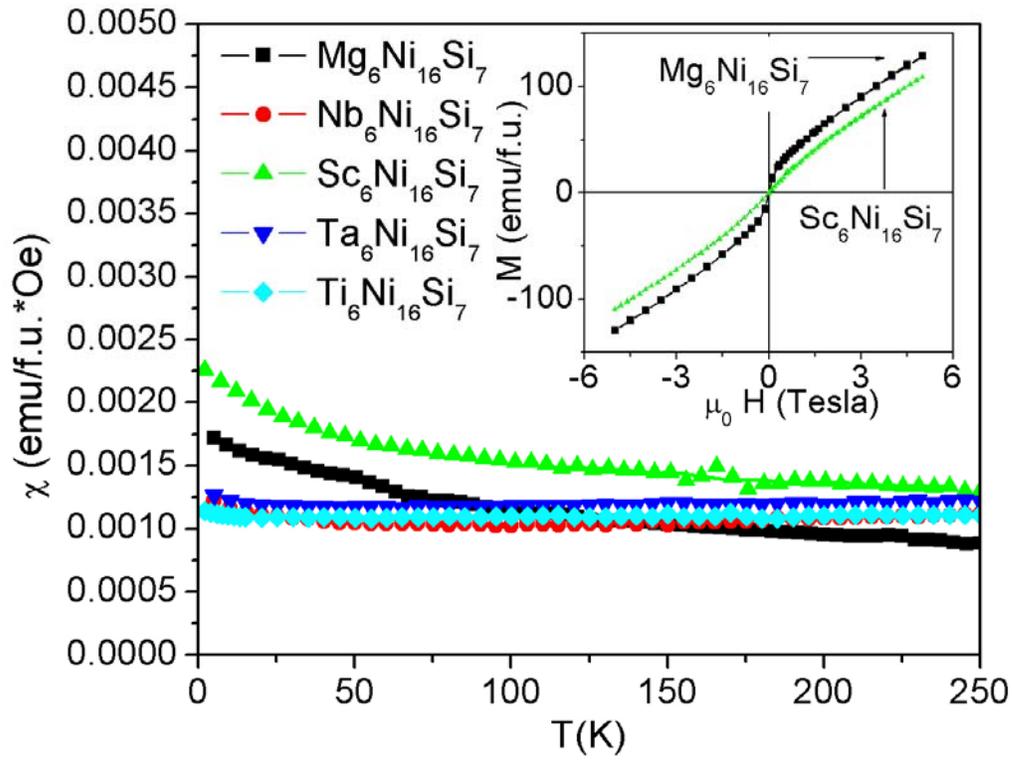

Figure **5**

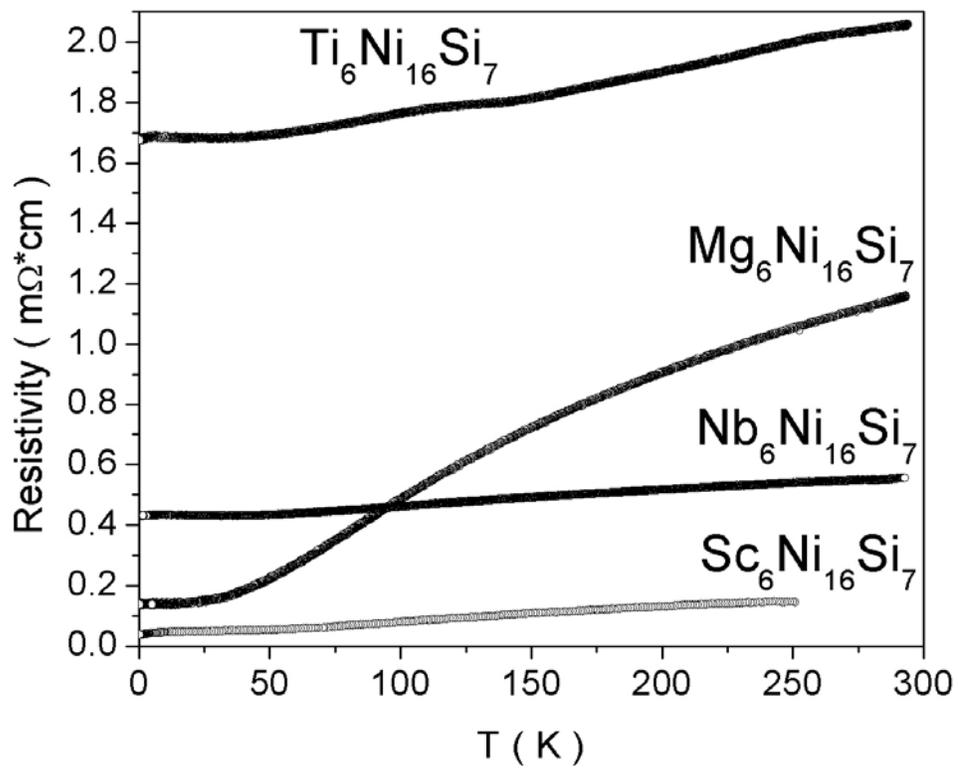

Figure **6**